\documentstyle[prc,aps,twocolumn]{revtex}
\draft

\newcommand{\ba}{\begin{eqnarray}}
\newcommand{\ea}{\end{eqnarray}}
\input{psfig}

\begin{document}

\title{Nuclear structure from random interactions}

\author{R. Bijker$^1$, A. Frank$^{1,2}$ and S. Pittel$^3$} 

\address{$^1$Instituto de Ciencias Nucleares, 
Universidad Nacional Aut\'onoma de M\'exico, \\
Apartado Postal 70-543, 04510 M\'exico, D.F., M\'exico \\ 
$^2$Centro de Ciencias F{\'{\i}}sicas, 
Universidad Nacional Aut\'onoma de M\'exico, \\
Apartado Postal 139-B, Cuernavaca, Morelos, M\'exico\\
$^3$ Bartol Research Institute, University of Delaware, 
Newark, Delaware 19716, U.S.A.}

\date{January 26, 2000}

\maketitle

\begin{abstract}
Low-lying states in nuclei are investigated using an ensemble of 
random interactions. Both in the nuclear shell model and in the 
interacting boson model we find a dominance of $J^P=0^+$ ground 
states. It is shown that this feature is not due to time reversal 
symmetry. In the shell model, evidence is found for the occurrence 
of pairing properties, and in the interacting boson model for 
both vibrational and rotational band structures. Our results 
suggest that these features represent general and robust properties 
of the model space, and do not depend on details of the interactions.  
\end{abstract} 


\section{Introduction}

Despite the fact that nuclei are complex many-body systems 
with many degrees of freedom, their spectral properties often 
show very regular features. 
A recent analysis of experimental energy systematics of medium and 
heavy even-even nuclei suggests a tripartite classification of 
nuclear structure into seniority, anharmonic vibrator and rotor 
regions \cite{Casten,Zamfir}. Plots of the excitation energies of the 
yrast states with $J^P=4^+$ against $J^P=2^+$ show a characteristic 
slope for each region: 1.00, 2.00 and 3.33, respectively. 
In each of these three regimes, the energy systematics is extremely 
robust. Moreover, the transitions 
between different regions occur very rapidly, typically with the 
addition or removal of only one or two pairs of nucleons. 
The transition between the seniority region (either semimagic or 
nearly semimagic nuclei) and the anharmonic vibrator regime (either 
vibrational or $\gamma$ soft nuclei) was addressed in a simple 
schematic shell model calculation and attributed to the proton-neutron 
interaction \cite{BFP}. The empirical characteristics of the 
collective regime which consists of the anharmonic vibrator and 
the rotor regions, as well as the transition between them,  
have been studied \cite{ZC,Jolos} in the framework of the 
interacting boson model (IBM) \cite{IBM}. An analysis of phase 
transitions in the IBM \cite{GK,DSI} has shown that the collective 
region is characterized by two basic phases (spherical and deformed) 
with a sharp transition region \cite{IZC,Dimitri}, rather than a gradual softening  
which is traditionally associated with the onset of deformation 
in nuclei.

In a separate development, the characteristics of low-energy 
spectra of many-body even-even nuclear systems have been studied 
recently in the context of the nuclear shell model (SM) with random 
two-body interactions \cite{JBD,Johnson}. Despite the random nature 
of the interactions, the low-lying spectra still show surprisingly 
regular features, such as a predominance of $J^P=0^+$ ground states 
separated by an energy gap from the excited states. This is contrary 
to the traditional wisdom in which the favoring of $J^P=0^+$ 
ground states is attributed to the nuclear pairing arising from the 
short-range nuclear force. A subsequent analysis of the pair 
transfer amplitudes has shown that pairing is a robust feature 
of the general two-body nature of shell model interactions and the 
structure of the model space \cite{JBDT}. On the other hand, no 
evidence was found for rotational band structures. 

The existence of robust features in the low-lying spectra 
of medium and heavy even-even nuclei \cite{Casten,Zamfir} suggests 
an underlying simplicity of low-energy nuclear 
structure never before appreciated. In order to address this point 
we carry out a study of the systematics of energy levels in the 
framework of the SM and the IBM with random interactions. We address 
time-reversal symmetry in connection with the dominance of $J^P=0^+$ 
ground states, look for regular spectral properties, and investigate 
the effect of many-body interactions. 

\section{The nuclear shell model}

We first consider the properties of nuclei in the presence of random
interactions within the context of the shell model.
As a model space we take that of $N$ identical nucleons in the $sd$ 
shell, which consists of single-particle orbitals with $j=1/2$, 3/2 
and 5/2~. The case of $N=6$ particles is one of the examples 
considered in \cite{JBD,Johnson} and referred to as corresponding to 
the nucleus $^{22}$O. 
For identical particles the isospin is the same for all states, 
and hence does not play a role. In Ref.~\cite{Johnson} it was shown 
that the single-particle energies have little effect on the results, 
and therefore they are not considered here. 
The two-body interactions can be expressed as 
\ba
H_2 &=& - \sum_{L=0}^4 \, \sum_{i \leq j} \zeta_{L_{ij}} \, 
(-1)^L \frac{P^{\dagger}_{L_i} \cdot \tilde{P}_{L_j} 
+ P^{\dagger}_{L_j} \cdot \tilde{P}_{L_i}}{1+\delta_{ij}} ~, 
\label{hsm2}
\ea
with
\ba
P^{\dagger}_{0_1} &=& 
(a^{\dagger}_{1/2} \times a^{\dagger}_{1/2})^{(0)}/\sqrt{2} ~, 
\nonumber\\
P^{\dagger}_{0_2} &=& 
(a^{\dagger}_{3/2} \times a^{\dagger}_{3/2})^{(0)}/\sqrt{2} ~, 
\nonumber\\
P^{\dagger}_{0_3} &=& 
(a^{\dagger}_{5/2} \times a^{\dagger}_{5/2})^{(0)}/\sqrt{2} ~, 
\nonumber\\
P^{\dagger}_{1_1} &=& 
(a^{\dagger}_{1/2} \times a^{\dagger}_{3/2})^{(1)} ~, 
\nonumber\\
P^{\dagger}_{1_2} &=& 
(a^{\dagger}_{3/2} \times a^{\dagger}_{5/2})^{(1)} ~, 
\nonumber\\
P^{\dagger}_{2_1} &=& 
(a^{\dagger}_{1/2} \times a^{\dagger}_{3/2})^{(2)} ~, 
\nonumber\\
P^{\dagger}_{2_2} &=& 
(a^{\dagger}_{1/2} \times a^{\dagger}_{5/2})^{(2)} ~, 
\nonumber\\
P^{\dagger}_{2_3} &=& 
(a^{\dagger}_{3/2} \times a^{\dagger}_{3/2})^{(2)}/\sqrt{2} ~, 
\nonumber\\
P^{\dagger}_{2_4} &=& 
(a^{\dagger}_{3/2} \times a^{\dagger}_{5/2})^{(2)} ~, 
\nonumber\\
P^{\dagger}_{2_5} &=& 
(a^{\dagger}_{5/2} \times a^{\dagger}_{5/2})^{(2)}/\sqrt{2} ~, 
\nonumber\\
P^{\dagger}_{3_1} &=& 
(a^{\dagger}_{1/2} \times a^{\dagger}_{5/2})^{(3)} ~, 
\nonumber\\
P^{\dagger}_{3_2} &=& 
(a^{\dagger}_{3/2} \times a^{\dagger}_{5/2})^{(3)} ~, 
\nonumber\\
P^{\dagger}_{4_1} &=& 
(a^{\dagger}_{3/2} \times a^{\dagger}_{5/2})^{(4)} ~, 
\nonumber\\
P^{\dagger}_{4_2} &=& 
(a^{\dagger}_{5/2} \times a^{\dagger}_{5/2})^{(4)}/\sqrt{2} ~. 
\ea
The coefficients $\zeta_{L_{ij}}$ correspond to the 30 two-body 
matrix elements for identical particles in the $sd$ shell. 
They are chosen independently from a Gaussian distribution of 
random numbers with zero mean and variance $v^2$, 
\ba
\left< \zeta_{L_{ij}} \zeta_{L'_{i'j'}} \right> &=& 
\delta_{LL'} \, (1+\delta_{ij,i'j'}) \, v^2 ~.
\label{goe}
\ea
Here $<>$ denotes an ensemble average. 
The ensemble of Eq.~(\ref{goe}) satisfies the requirement that 
it is invariant under orthogonal transformations 
({\it i.e.} a change of basis). The variance of the Gaussian 
distribution $v^2$ is independent of the angular momentum and 
only represents an overall energy scale. All results that we present
here were obtained using 1000 runs.

\begin{figure}[tbp]
\centerline{\hbox{
\psfig{figure=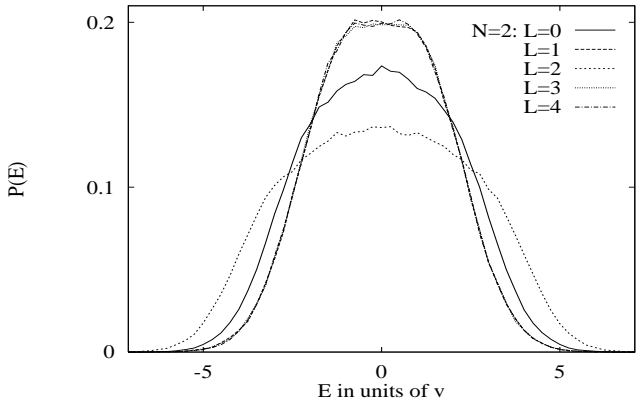,height=0.3\textwidth,width=0.5\textwidth} }}
\vspace{15pt}
\caption[]{Level distributions for $N=2$ particles ($^{18}$O).}
\label{sd2}
\end{figure}

For $N=2$ particles the Hamiltonian 
matrix is entirely random and the ensemble coincides with the 
Gaussian Orthogonal Ensemble (GOE), which is characterized by 
a semi-circular distribution of eigenvalues \cite{Wigner}
\ba
P(E) &=& \frac{1}{2 \pi d v^2} \sqrt{4dv^2-E^2} ~,
\label{pe1}
\ea
whose width depends on the dimension $d$ of the Hamiltonian 
matrix \cite{Brody} according to
\ba
\sigma &=& \sqrt{ \frac{\left< \mbox{Tr} H^2 \right>}{d} - 
\left( \frac{\left< \mbox{Tr} H \right>}{d} \right)^2 } 
\;=\; \sqrt{(d+1)v^2} ~. 
\label{width}
\ea
In Table~\ref{O18} we show the percentage of the total number 
of runs for which the ground state has a given angular momentum. 
Clearly, the angular momentum for which the width of the distribution
is largest will be the most likely to be the ground state. In this case,
as noted earlier, the widths depend directly on the corresponding dimension
of the basis. 
Thus, the $J=2$ state is the most likely to be 
the ground state, followed by $J=0$ and then by $J=1,3,4$, exactly
as seen in the table. These various points are made clearer from the 
semi-circular level distributions shown in Fig.~\ref{sd2}.  

\begin{figure}[tbp]
\centerline{\hbox{
\psfig{figure=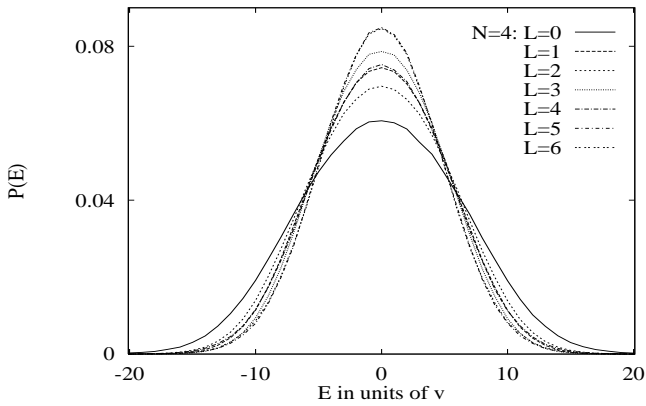,height=0.3\textwidth,width=0.5\textwidth} }}
\vspace{15pt}
\caption[]{Level distributions for $N=4$ particles ($^{20}$O).}
\label{sd4}
\end{figure}

\begin{figure}[tbp]
\centerline{\hbox{
\psfig{figure=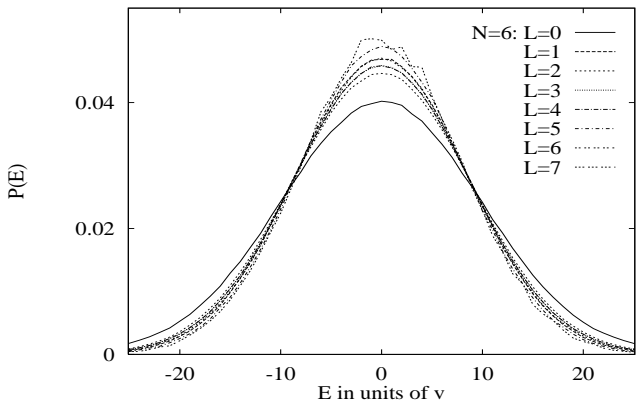,height=0.3\textwidth,width=0.5\textwidth} }}
\vspace{15pt}
\caption[]{Level distributions for $N=6$ particles ($^{22}$O).}
\label{sd6}
\end{figure}

For $N>2$ particles the ensemble 
is the Two-Body Random Ensemble (TBRE) \cite{French,Bohigas}, 
in which the $N$-body matrix elements are correlated and can 
be expressed in terms of the random two-body matrix elements 
of Eq.~(\ref{goe}) by the usual reduction formulae. 
The eigenvalues now follow a Gaussian distribution 
\cite{French,Bohigas,Gervois}
\ba
P(E) &=& \frac{1}{\sigma\sqrt{2\pi}} \mbox{e}^{-E^2/2\sigma^2} ~. 
\label{pe2}
\ea
As an example, we consider the nuclei $^{20,22}$O with four 
and six valence neutrons. 
In Table~\ref{oxygen} we show the percentage of 
the total number of runs for which the ground state has a given 
angular momentum. The corresponding level distributions are shown
in Figs.~\ref{sd4} and \ref{sd6}.  
For $N=4$ particles the percentage 
of $J^P=0^+$ ground states is 55.9 $\%$, significantly larger than for
$N=2$ (see Table~\ref{O18}). Note, however, that the percentage of 
$0^+$ states in the model space is only 11.1 $\%$ for this case. 
Similar results hold 
for $N=6$ particles (see the last column of Table~\ref{oxygen}); namely 
the percentage of $0^+$ ground states dramatically exceeds the
percentage of $0^+$ states in the basis. 
Our results for $N=6$ are in agreement with those obtained earlier by 
Johnson et al. \cite{JBD,Johnson}. 

In the cases of $N=4$ and $6$, the percentage of ground states associated
with each angular momentum is also correlated with the widths 
of the distributions, which are now Gaussian (see Figs.~\ref{sd4} 
and \ref{sd6}). The key difference 
between these results and those for 
$N=2$ is that here there is no direct connection between the width and the
size of the basis for a given angular momentum. $J^P=0^+$ ground 
states predominate for $N=4$ and $6$ even though they
have much fewer basis states than some of the other angular 
momenta.

The observed preponderance of $J^P=0^+$ ground states for $N>2$ is 
surprising, considering that there is no obvious pairing 
character in the assumed random forces. 
Thus the question remains: what is it 
that produces this dominance of $0^+$ ground states in even-even 
many-body systems? One possibility is that it arises 
because of the time-reversal 
invariance of the random Hamiltonian. Since time-reversed states 
play an important role in the formation of correlated $0^+$ 
(Cooper) pairs which in turn can give rise to favored collective 
many-body states, it is conceivable that time-reversal invariance 
may contain a built-in preference for $J^P=0^+$ many-body ground 
states \cite{random}.

To see whether this is indeed the case, we consider what
happens when we break time-reversal 
invariance in the random two-body interactions. This can be done 
by introducing a Gaussian Unitary Ensemble (GUE) rather than a 
Gaussian Orthogonal Ensemble (GOE) to randomly generate the 
two-body matrix elements. More specifically, we consider a 
two-body Hamiltonian of the form \cite{Brody,FKPT} 
\ba
H_2 &=& - \sum_{L=0}^4 \, \sum_{i \leq j} \left[ \zeta_{L_{ij}} \, 
\frac{P^{\dagger}_{L_i} \cdot \tilde{P}_{L_j} 
+ P^{\dagger}_{L_j} \cdot \tilde{P}_{L_i}}{1+\delta_{ij}} \right.
\nonumber\\
&& \left. + i \epsilon \eta_{L_{ij}} \, 
\frac{P^{\dagger}_{L_i} \cdot \tilde{P}_{L_j} 
- P^{\dagger}_{L_j} \cdot \tilde{P}_{L_i}}{1+\delta_{ij}} \right] 
\frac{(-1)^L}{\sqrt{1+\epsilon^2}} ~.  
\label{tr}
\ea
The coefficients $\zeta_{L_{ij}}$ and $\eta_{L_{ij}}$ are 
chosen independently from a Gaussian distribution of random 
numbers with zero mean and variance $v^2$ as 
\ba
\left< \zeta_{L_{ij}} \zeta_{L'_{i'j'}} \right> &=& 
\delta_{LL'} \, (1+\delta_{ij,i'j'}) \, v^2 ~,
\nonumber\\
\left< \eta_{L_{ij}} \eta_{L'_{i'j'}} \right> &=& 
\delta_{LL'} \, (1-\delta_{ij,i'j'}) \, v^2 ~,
\nonumber\\
\left< \zeta_{L_{ij}} \eta_{L'_{i'j'}} \right> &=& 0 ~. 
\label{gue}
\ea
For $\epsilon=0$ and 1 they 
correspond to GOE and GUE, respectively. The Hamiltonian is 
time-reversal invariant if the two-body matrix elements are real, 
{\it i.e.} $\epsilon=0$. The breaking of time-reversal symmetry can 
be studied by taking $0 \leq \epsilon \leq 1$. For a given value of 
$J$, the above ensemble for two-body interactions gives a 
semicircle level density. The normalization is chosen such that the 
radius of this semicircle distribution does not depend on $\epsilon$ 
\cite{FKPT}. In Table~\ref{time} we present the results for $N=6$ 
identical particles in the $sd$ shell. For $0 < \epsilon \leq 1$ 
the time-reversal invariance is broken. We see that the dominance of 
$0^+$ ground states {\em increases} with $\epsilon$ from 
67.7 to 76.8 $\%$. On the basis of these results, we 
conclude that time-reversal invariance of the two-body interactions 
is not the origin of the dominance of $0^+$ ground states. 

\begin{figure}
\centerline{\hbox{
\psfig{figure=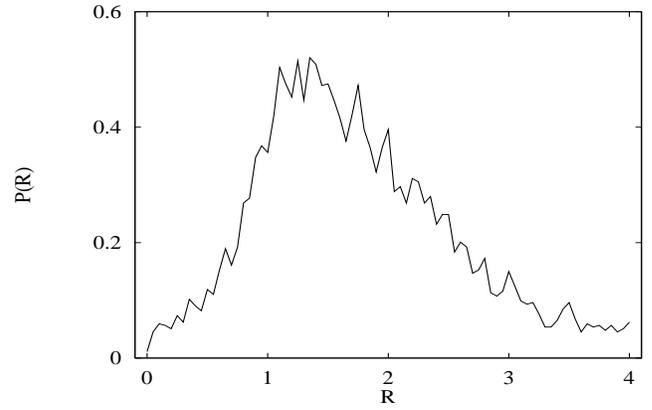,height=0.3\textwidth,width=0.5\textwidth} }}
\vspace{15pt}
\caption[]{Probability distribution $P(R)$ of the energy ratio 
$R=[E(4_1^+)-E(0_1^+)]/[E(2_1^+)-E(0_1^+)]$ with $\int P(R) dR = 1$ 
for $N=6$ particles in the $sd$ shell with random two-body 
interactions.}
\label{sm}
\end{figure}

For the cases with a $J^P=0^+$ ground state we calculate the 
probability distribution of the energy ratio 
\ba
R &=& \frac{E(4_1^+)-E(0_1^+)}{E(2_1^+)-E(0_1^+)} ~. 
\label{e4e2}
\ea
As noted earlier, this energy ratio has  
characteristic values of $R\approx1$, $2$ and $10/3$ for the seniority, 
vibrational and rotational regions, respectively. 
In Fig.~\ref{sm} we show the results for $N=6$ particles. The 
probability distribution shows a broad peak 
between $1 \leq R \leq 2$, with a maximum around $1.3$. This suggests that
on average a system of identical nucleons tends to behave in accord with
the seniority regime of \cite{Casten,Zamfir}.
An analysis of the amplitudes for pair transfer 
between ground states  
has shown that pairing is a robust feature of two-body shell model 
interactions and arises from a much broader class of Hamiltonians 
than the ones usually considered \cite{JBDT}. This finding and ours 
are in qualitative agreement with the empirical observation of very 
robust spectroscopic features in the seniority regime 
\cite{Casten,Zamfir}.
Since the distribution extends to $R=2$, we conclude that there is 
also some evidence, although minimal, for vibrational structure 
in our calculations. On the other hand, there is no evidence 
for the occurrence of rotational bands, at least within the context of the 
model space we have considered (see also \cite{JBD}). 

In the next section we carry out a study of the systematics 
of collective levels in the IBM with random interactions, and look 
for evidence for vibrational and rotational bands. 

\section{The interacting boson model}

In the IBM collective nuclei are described as a system of $N$ 
interacting monopole and quadrupole bosons \cite{IBM}.
The one-body Hamiltonian of the model contains the single-boson energies 
\ba 
H_1 &=& \epsilon_0 \, s^{\dagger} \cdot \tilde{s}  
+ \epsilon_2 \, d^{\dagger} \cdot \tilde{d} ~, 
\label{h1}
\ea
and the two-body Hamiltonian contains the various two-boson interactions 
\ba
H_2 &=& \sum_{L=0,2,4} \, \sum_{i \leq j} \zeta_{L_{ij}} \, 
\frac{P^{\dagger}_{L_i} \cdot \tilde{P}_{L_j} 
+ P^{\dagger}_{L_j} \cdot \tilde{P}_{L_i}}{1+\delta_{ij}} ~, 
\label{h2}
\ea
with
\ba
P^{\dagger}_{0_1} &=& 
(s^{\dagger} \times s^{\dagger})^{(0)}/\sqrt{2} ~, 
\nonumber\\
P^{\dagger}_{0_2} &=& 
(d^{\dagger} \times d^{\dagger})^{(0)}/\sqrt{2} ~, 
\nonumber\\
P^{\dagger}_{2_1} &=& 
(s^{\dagger} \times d^{\dagger})^{(2)} ~, 
\nonumber\\
P^{\dagger}_{2_2} &=& 
(d^{\dagger} \times d^{\dagger})^{(2)}/\sqrt{2} ~, 
\nonumber\\
P^{\dagger}_{4_1} &=& 
(d^{\dagger} \times d^{\dagger})^{(4)}/\sqrt{2} ~.
\ea
The coefficients $\epsilon_L$ and $\zeta_{L_{ij}}$ 
correspond to the 2 one-body and 7 two-body 
matrix elements, respectively. They are chosen 
independently from a Gaussian distribution of random numbers  
with zero mean and variance $v^2$ according to 
\ba
\left< \epsilon_L \epsilon_{L'} \right> &=& 
\delta_{LL'} \, 2 \, v^2 ~, 
\nonumber\\
\left< \zeta_{L_{ij}} \zeta_{L'_{i'j'}} \right> &=& 
\delta_{LL'} \, (1+\delta_{ij,i'j'}) \, v^2 ~.
\label{ensemble2}
\ea
First we consider the most general one- and two-body IBM 
Hamiltonian
\ba
H_{12} &=& \frac{1}{N} \left[ H_1 + \frac{1}{N-1} H_2 \right] ~.
\label{h12}
\ea
Note that to remove the $N$ dependence of the matrix elements of 
$k$-body interactions, we have scaled $H_k$ by 
$\prod_{i=1}^k (N+1-i)$. 
In all calculations we take $N=16$ and 1000 runs. For each set of 
randomly generated many-body matrix elements we calculate the entire 
energy spectrum and the $B(E2)$ values between the yrast states. 

Just as in the case of the nuclear shell model \cite{JBD}, we 
find a predominance of $J^P=0^+$ ground states; 63.4 $\%$
of the ground states have this value of the angular momentum, even 
though only 3.3 $\%$ of the basis states do. Other angular 
momenta for which there are relatively high ground-state probabilities 
are $J^P=2^+$ (13.8 $\%$) and $J^P=32^+$ -- the maximum value of the 
angular momentum (16.7 $\%$). 

\begin{figure}
\centerline{\hbox{
\psfig{figure=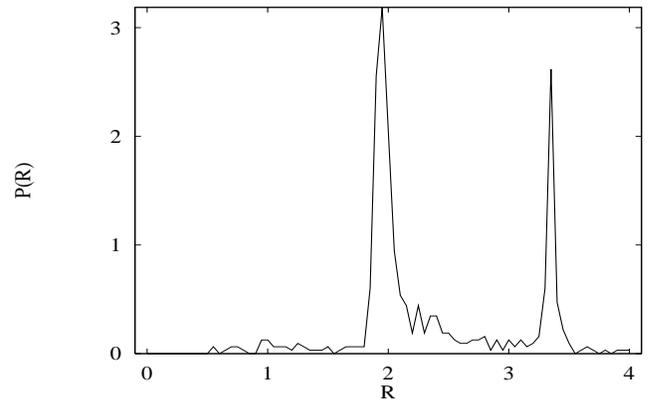,height=0.3\textwidth,width=0.5\textwidth} }}
\vspace{15pt}
\caption[]{Probability distribution $P(R)$ of the energy ratio 
$R=[E(4_1^+)-E(0_1^+)]/[E(2_1^+)-E(0_1^+)]$ with $\int P(R) dR = 1$ 
in the IBM with random one- and two-body interactions. 
The number of bosons is $N=16$.}
\label{ratio12}
\end{figure}

For those cases having a $J^P=0^+$ ground state 
we have calculated the probability distribution of the 
energy ratio $R$ of Eq.~(\ref{e4e2}). 
Fig.~\ref{ratio12} shows a remarkable result: the probability 
distribution $P(R)$ has two very pronounced peaks, one at 
$R \sim 1.95$ and a narrower one at $R \sim 3.35$ \cite{BF}. 
These values correspond almost exactly to the harmonic vibrator 
and rotor values of 2 and 10/3 (see Table~\ref{BE2}). 

\begin{figure}
\centerline{\hbox{
\psfig{figure=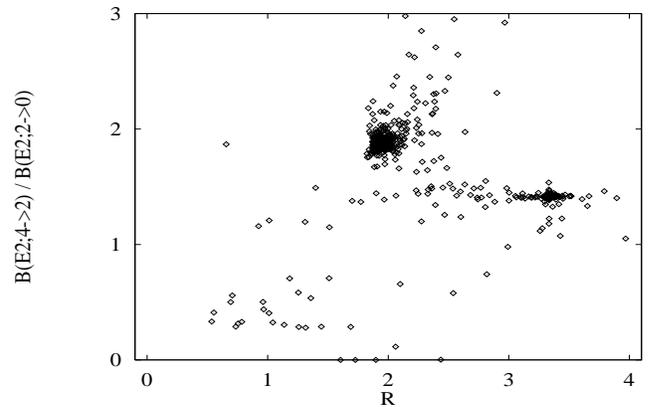,height=0.3\textwidth,width=0.5\textwidth} }}
\vspace{15pt}
\caption[]{Correlation between ratios of $B(E2)$ values and energies 
in the IBM with random one- and two-body interactions. 
The number of bosons is $N=16$.}
\label{corr}
\end{figure}

Energies by themselves are not sufficient to decide whether 
or not there exists a collective structure. Levels belonging to a 
collective band are connected by strong electromagnetic 
transitions. In Fig.~\ref{corr} we show a correlation plot between 
the ratio of $B(E2)$ values for the $4_1^+ \rightarrow 2_1^+$ and 
$2_1^+ \rightarrow 0_1^+$ transitions and the energy ratio R. For  
the $B(E2)$ values we use the quadrupole operator
\ba
\hat Q_{\mu}(\chi) &=& ( s^{\dagger} \tilde{d} 
+ d^{\dagger} s)^{(2)}_{\mu} 
+ \chi \, (d^{\dagger} \tilde{d})^{(2)}_{\mu} ~, 
\label{qop}
\ea
with $\chi=-\sqrt{7}/2$. 
For completeness, in Table~\ref{BE2} we show the results for the 
three symmetry limits of the IBM \cite{IBM}. In the large $N$ limit, 
the ratio of $B(E2)$ values is 2 for the harmonic vibrator 
and 10/7 both for the $\gamma$-unstable rotor and the pure rotor. 
There is a strong correlation 
between the first peak in the energy ratio and the vibrator value 
for the ratio of $B(E2)$ values (the concentration of points in this 
region corresponds to about 50 $\%$ of all cases), and for 
the second peak and the rotor value (about 25 $\%$ of all cases) 
\cite{BF}. 

\begin{figure}
\centerline{\hbox{
\psfig{figure=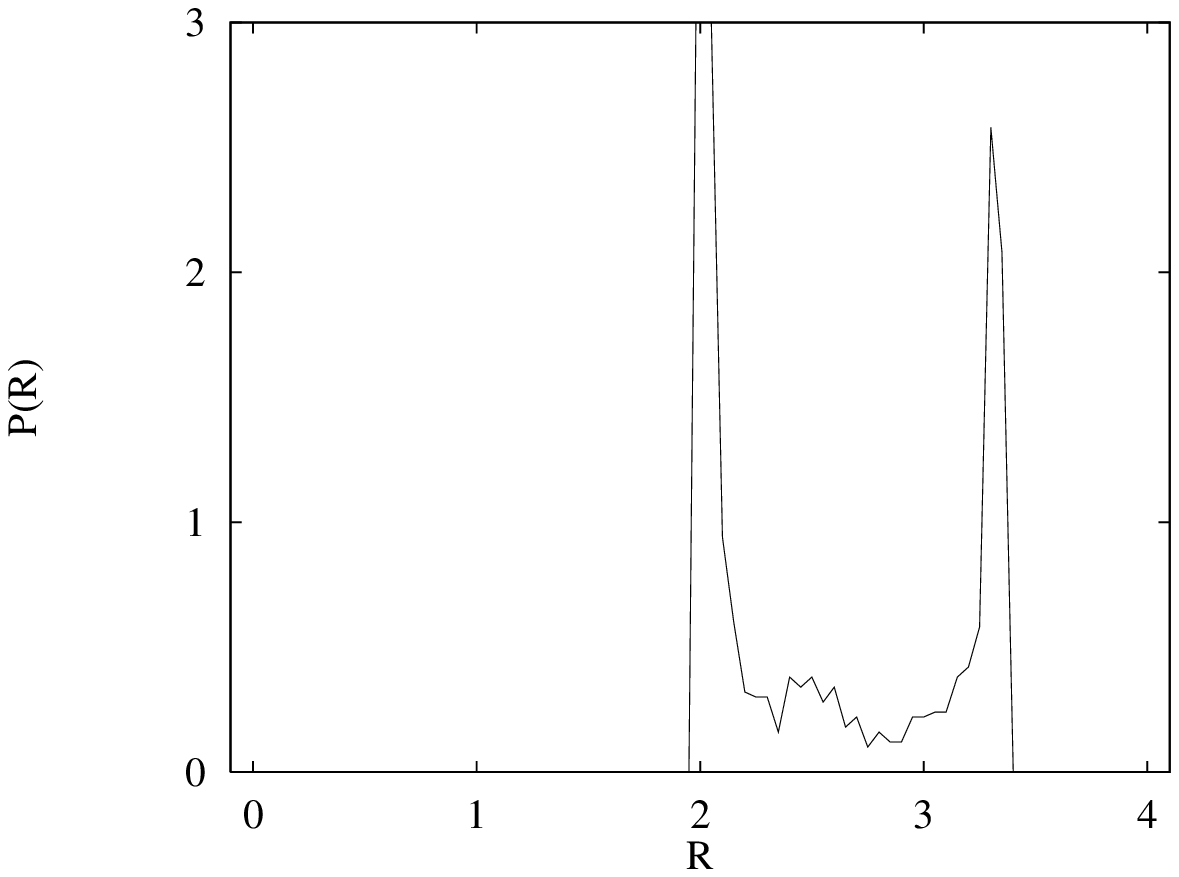,height=0.3\textwidth,width=0.5\textwidth} }}
\vspace{15pt}
\caption[]{As Fig.~\protect\ref{ratio12}, but 
in the consistent-Q formulation of the IBM.}
\label{cqf1}
\end{figure}

\begin{figure}
\centerline{\hbox{
\psfig{figure=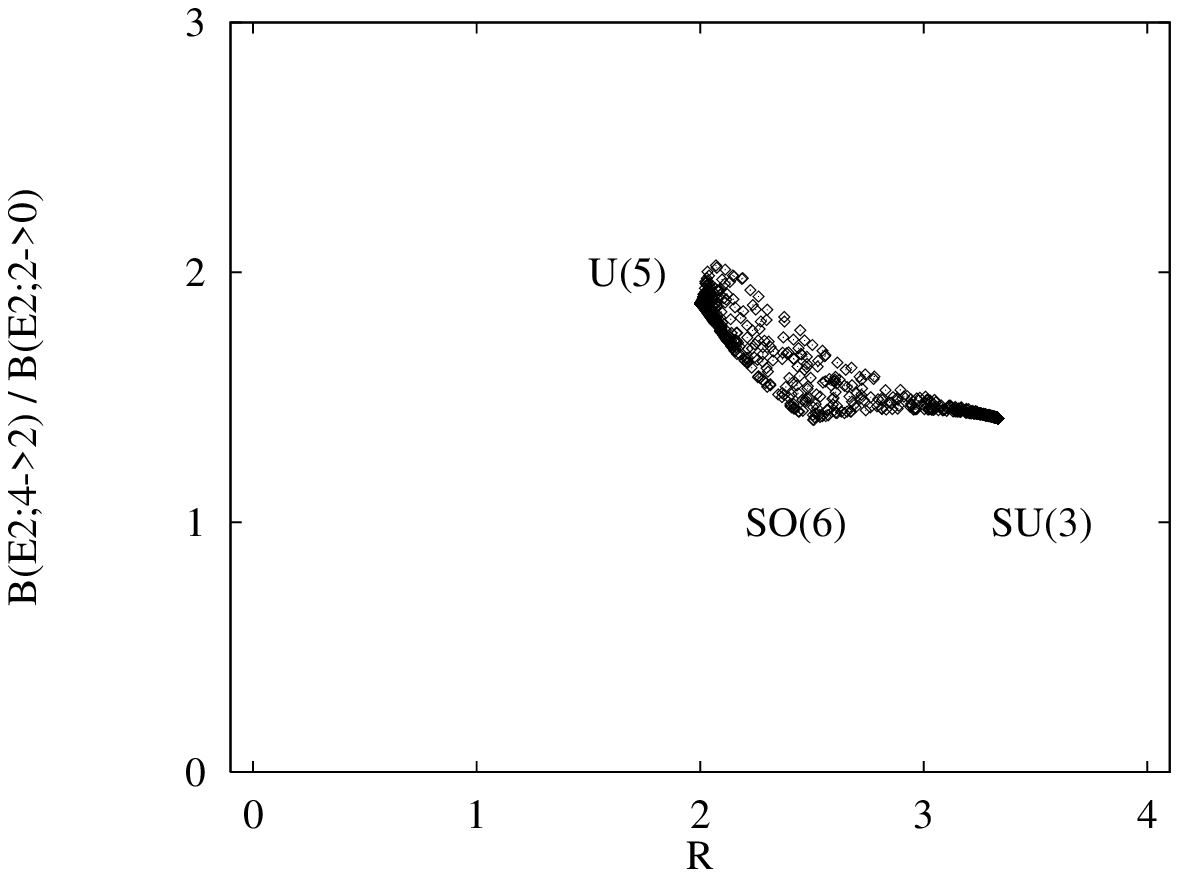,height=0.3\textwidth,width=0.5\textwidth} }}
\vspace{15pt}
\caption[]{As Fig.~\protect\ref{corr}, but 
in the consistent-Q formulation of the IBM.}
\label{cqf2}
\end{figure}

The results presented in Figs.~\ref{ratio12} and \ref{corr} were 
obtained with random interactions, with no restriction on the 
sign nor the magnitude of the one- and two-body matrix elements. 
It is of interest to make a comparison with a calculation in 
which the parameters are restricted to the `physically' 
allowed region. To this end, we consider the consistent-Q 
formulation \cite{CQF} which uses the same form for the 
quadrupole operator, Eq.~(\ref{qop}), {\it i.e.} with the same 
value of $\chi$ for the $E2$ operator and the Hamiltonian 
\ba
H &=& \epsilon \, \hat n_d 
- \kappa \, \hat Q(\chi) \cdot \hat Q(\chi) ~. 
\label{hcqf}
\ea
The parameters $\epsilon$ and $\kappa$ are restricted to be positive, 
whereas $\chi$ can be either positive or negative 
$-\sqrt{7}/2 \leq \chi \leq \sqrt{7}/2$. The properties of the 
Hamiltonian of Eq.~(\ref{hcqf}) are investigated by taking 
the scaled parameters $\eta=\epsilon/[\epsilon+4\kappa(N-1)]$ and 
$\bar{\chi}=-2\chi/\sqrt{7}$ randomly on the intervals 
$0 \leq \eta \leq 1$ and $-1 \leq \bar{\chi} \leq 1$ (these 
coefficients have been used as control parameters in a study of 
phase transitions in the IBM \cite{DSI,IZC}).  
In Figs.~\ref{cqf1} and \ref{cqf2} we show the corresponding 
probability distribution and correlation plot 
for the consistent-Q formulation of the IBM 
with realistic interactions. Although in this case the points 
are concentrated in a smaller region of the plot than before, 
the results show the same qualititative behavior as  
with random one- and two-body interactions. In Fig.~\ref{cqf2} 
we have identified each of the dynamical symmetries 
of the IBM (and the transitions between them). 
There is a large overlap between the regions with the highest 
concentration of points in Figs.~\ref{corr} and \ref{cqf2}. 

These results, {\it i.e.} the dominance of $J^P=0^+$ ground 
states and the occurrence of both vibrational and rotational 
structures, are not based solely on energies, but also involve 
wave function information via the quadrupole transitions. The use of 
random interactions (both in magnitude and sign) show that these 
regular features arise from a much wider class of Hamiltonians 
than are generally considered to be realisitic, and are,  
to a certain extent, independent of the specific character 
of the interactions. This too is in qualitative agreement with 
the empirical 
observation of robust features in the low-lying spectra of 
medium and heavy even-even nuclei \cite{Casten,Zamfir}.  
This leads naturally to the question of what is the cause of this 
behavior, if the only ingredients of the calculations are the 
one- and two-body nature of the interactions, the number of bosons 
$N$ and the structure of the model space? 

\begin{figure}
\centerline{\hbox{
\psfig{figure=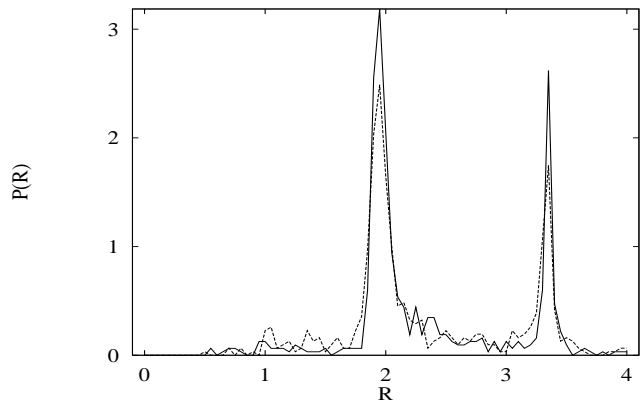,height=0.3\textwidth,width=0.5\textwidth}}}
\vspace{15pt}
\caption[]{As Fig.~\protect\ref{ratio12}, but for random one- and 
two-body interactions $H_{12}$ (solid line) and random one-, two- 
and three-body interactions $H_{123}$ (dashed line).}
\label{ratio123}
\end{figure}

In order to see to what extent the results found above \cite{BF} 
depend on the rank of the interactions, we study the effect of 
the inclusion of three-body interactions in the Hamiltonian. 
Three-body interactions are of special interest in the IBM, 
since they can give rise to stable triaxial deformations 
\cite{Piet}, which are absent in the case of one- and two-body 
interactions only. We consider the most general one-, two- 
and three-body IBM Hamiltonian
\ba
H_{123} &=& \frac{1}{N} \left[ H_1 + \frac{1}{N-1} \left[ H_2  
+ \frac{1}{N-2} H_3 \right] \right] ~,
\label{h123}
\ea
where $H_1$ and $H_2$ are given in Eqs.~(\ref{h1}) and (\ref{h2}). 
$H_3$ contains the 17 possible three-body random interactions and 
can be written in a similar fashion. 
Again we find a dominance (61.9 $\%$) of $J^P=0^+$ ground states; 
in 12.0 $\%$ of the cases the ground state has $J^P=2^+$, 
and in 17.7 $\%$ it has the maximum value of the angular momentum 
$J^P=32^+$, very close to the results for the case of one- and 
two-body interactions only (63.4, 13.8 and 16.7 $\%$, respectively). 
Also the probability distribution $P(R)$ shows the same behavior 
as for one- and two-body interactions. 
In Fig.~\ref{ratio123} we compare the result for $H_{12}$ (solid curve, 
see Fig.~\ref{ratio12}) with $H_{123}$. In both cases we see a 
very clear structure with pronounced peaks at the vibrational 
and rotational values of the energy ratio. Since the Hamiltonian 
$H_{123}$ depends on 26 independent random numbers 
(2 one-body, 7 two-body and 17 three-body matrix elements),  
there is less correlation between its many-body ($N=16$) matrix 
elements than in the case of $H_{12}$. This results in somewhat 
less pronounced (lower and broader) -- but nevertheless very clear -- 
peaks in the probability distribution $P(R)$. 

\section{Summary and conclusions}

In this work, we discussed global properties of nuclear structure 
using random Hamiltonians.

We first considered the problem from a shell model perspective, 
focussing on a system of identical neutrons in the $sd$ shell.
We confirmed the conclusion reached earlier by Johnson and 
collaborators \cite{JBD,Johnson} that nuclei with many nucleons 
favor $J^P=0^+$ ground states, even without a dominant
pairing component in the force. We demonstrated further that 
this is not a consequence of time-reversal invariance of the 
random Hamiltonian. Finally, we showed that systems of identical 
nucleons interacting via random two-body interactions 
tend to favor a seniority structure, 
in accord with conclusions reported recently in \cite{JBDT}.
There is little or no evidence for the occurrence of vibrational 
and rotational bands. 

We then considered the same issues in the context of the IBM, a 
collective model that from the outset emphasizes the role of monopole
and quadrupole pairs. Here too we found that $0^+$ ground states 
predominate, exactly as in the shell model analysis. In contrast, we
found that the IBM strongly favors both vibrational and rotational
structures, as evident from energy ratios of low-lying states
and their corresponding $BE(2)$ ratios. These conclusions emerged 
from a much wider class of Hamiltonians than is usually thought 
to be `realistic'. The inclusion of three-body random interactions 
did not change these basic conclusions, as long as the number of 
bosons is sufficiently large. This suggests 
that the observed vibrational and rotational features represent 
general and robust properties of the IBM model space, and do not  
depend significantly on details of the interaction. Since the structure 
of the model space is completely determined by the degrees of 
freedom, our results emphasize once again the importance 
of the selection of the relevant degrees of freedom. 

The results that we obtained with random Hamiltonians
in the shell model and the IBM are in qualitative agreement 
with the empirical observation of robust features in the low-lying 
spectra of medium and heavy even-even nuclei and their tripartite
classification into seniority, anharmonic vibrator and rotor 
regimes \cite{Casten,Zamfir}. The analysis with random interactions 
shows that seniority arises as a global property of the shell model 
space, while vibrational and rotational bands arise as general 
features of the interacting boson model space. 
However, the IBM is based on the assumption that low-lying collective 
excitations in nuclei can be described as a system of interacting 
monopole and quadrupole bosons, which in turn are associated with 
generalized pairs of like-nucleons with angular momentum $L=0$ 
and $L=2$. Therefore it would be very important to establish 
whether vibrational and rotational 
features can also arise from ensembles of random interactions in 
the nuclear shell model, if appropriate (minimal) restrictions 
are imposed on the parameter space. 

\section*{Acknowledgements}

It is a pleasure to thank Rick Casten, Jorge Flores 
and Franco Iachello for illuminating discussions. This work was 
supported in part by DGAPA-UNAM under project IN101997, 
by CONACyT under projects 32416-E and 32397-E, 
and by NSF under grant PHY-9970749.

\clearpage

\begin{table}[h]
\centering
\caption[]{Percentage of ground states with angular momentum $J$ for 
GOE with $N=2$ identical particles in the $sd$ shell 
(the nucleus $^{18}$O). }
\label{O18} 
\vspace{15pt}
\begin{tabular}{ccrrrr}
& & & & & \\
$N$ & $J$ & $d$ & $\sigma$ & Basis & GOE \\
& & & & & \\
\hline
& & & & & \\
2 & 0 & 3 & 2.00 & 21.4 $\%$ & 15.9 $\%$ \\
  & 1 & 2 & 1.73 & 14.3 $\%$ &  4.9 $\%$ \\
  & 2 & 5 & 2.45 & 35.7 $\%$ & 68.3 $\%$ \\
  & 3 & 2 & 1.73 & 14.3 $\%$ &  6.1 $\%$ \\
  & 4 & 2 & 1.74 & 14.3 $\%$ &  4.8 $\%$ \\
& & & & & \\
\end{tabular}
\end{table}

\begin{table}
\centering
\caption[]{Percentage of ground states with angular momentum $J$ 
for TBRE with $N=4$ and 6 identical particles in the $sd$ shell 
(the nuclei $^{20,22}$O). }
\label{oxygen} 
\vspace{15pt}
\begin{tabular}{ccrrrr}
& & & & & \\
$N$ & $J$ & $d$ & $\sigma$ & Basis & TBRE \\
& & & & & \\
\hline
& & & & & \\
4 & 0 &  9 & 6.24 & 11.1 $\%$ & 55.9 $\%$ \\
  & 1 & 12 & 5.05 & 14.8 $\%$ &  4.9 $\%$ \\
  & 2 & 21 & 5.37 & 25.9 $\%$ & 22.7 $\%$ \\
  & 3 & 15 & 4.79 & 18.5 $\%$ &  1.4 $\%$ \\
  & 4 & 15 & 5.12 & 18.5 $\%$ & 12.3 $\%$ \\
  & 5 &  6 & 4.65 &  7.4 $\%$ &  1.5 $\%$ \\
  & 6 &  3 & 4.69 &  3.7 $\%$ &  1.3 $\%$ \\
& & & & & \\
\hline
& & & & & \\
6 & 0 & 14 & 10.16 &  9.9 $\%$ & 67.7 $\%$ \\
  & 1 & 19 &  8.53 & 13.4 $\%$ &  1.3 $\%$ \\
  & 2 & 33 &  9.01 & 23.2 $\%$ & 15.0 $\%$ \\
  & 3 & 29 &  8.80 & 20.4 $\%$ &  7.1 $\%$ \\
  & 4 & 26 &  8.80 & 18.3 $\%$ &  6.8 $\%$ \\
  & 5 & 12 &  8.27 & 8.5 $\%$ &  0.4 $\%$ \\
  & 6 &  8 &  8.61 & 5.6 $\%$ &  1.7 $\%$ \\
  & 7 &  1 &  8.07 & 0.7 $\%$ &  0.0 $\%$ \\
& & & & \\
\end{tabular}
\end{table}

\begin{table}
\caption[]{Percentage of $J^P=0^+$ ground states for TBRE with 
$N=6$ identical particles in the $sd$ shell (the nucleus $^{22}$O) 
as a function of $\epsilon$. }
\label{time}
\vspace{15pt}
\begin{tabular}{cc}
& \\
$\epsilon$ & TBRE \\
& \\
\hline
& \\
0.00 & 67.7 $\%$ \\
0.25 & 69.3 $\%$ \\
0.50 & 71.7 $\%$ \\
0.75 & 74.0 $\%$ \\
1.00 & 76.8 $\%$ \\
& \\
\end{tabular}
\end{table}

\begin{table}
\caption[]{Energies and $B(E2)$ values in the dynamical symmetry 
limits of the IBM \protect\cite{IBM}. In the $U(5)$ and $SO(6)$ 
limits we show the result for the leading order contribution to the 
rotational spectra.} 
\label{BE2}
\vspace{15pt}
\begin{tabular}{ccc}
& \\
& $\frac{E(4_1^+)-E(0_1^+)}{E(2_1^+)-E(0_1^+)}$  
& $\frac{B(E2;4_1^+ \rightarrow 2_1^+)}{B(E2;2_1^+ \rightarrow 0_1^+)}$ \\
& \\
\hline
& \\
$U(5)$  & $2$ & $\frac{2(N-1)}{N}$ \\
$SO(6)$ & $\frac{5}{2}$ 
& $\frac{10(N-1)(N+5)}{7N(N+4)}$ \\
$SU(3)$ & $\frac{10}{3}$ 
& $\frac{10(N-1)(2N+5)}{7N(2N+3)}$ \\
& \\
\end{tabular}
\end{table}

\end{document}